\documentclass[12pt]{article}
\usepackage[english]{babel}
\usepackage{graphicx}

\begin{document}

\begin{center}
\begin{large}
\textbf{ONCE MORE ON THE DERIVATION OF THE DIRAC EQUATION}
\end{large}
\vskip 0.5cm

\textbf{V.M. Simulik$^1$, I.Yu. Krivsky$^1$}
\end{center}

\begin{center}
\textit{$^1$Institute of Electron Physics, National Academy of
Sciences of Ukraine, 21 Universitetska Str., 88000 Uzhgorod,
Ukraine; E-mail: vsimulik@gmail.com}

\end{center}

\vskip 1.cm

\noindent ABSTRACT. A brief review of the different ways of the Dirac equation derivation is given. The foundations of the relativistic canonical quantum mechanics of a fermionic doublet are formulated. In our approach the Dirac equation is derived from the main equation of relativistic canonical quantum mechanics. In the formalism of relativistic canonical quantum mechanics the absence of necessity to appeal for the conception of the negative mass of the antiparticle is shown.

\vskip 0.5cm


\vskip 0.5cm


\vskip 1.cm

\section{Introduction}

It may seem from the title of the article that the subject of our investigation is trivial. It may seem for everybody that he knows the quantum-mechanical description of spin $s=\frac{1}{2}$ doublet! Indeed, the old imagination that Dirac equation gives the relativistic quantum-mechanical description of the fermionic doublet still is widespread.

Nevertheless, in some principal quantum-mechanical problems such description is not satisfactory. L. Foldy and S. Wouthuysen [1] suggested the nonlocal canonical formulation of equation for the spinor field. In this description [1--3], first of all for the electron having spin $\overrightarrow{s}=\frac{1}{2}\overrightarrow{\sigma}$, many quantum-mechanical details were clarified. The   equation $i\partial_{t}f(x)=\sqrt{m^{2}-\Delta}f(x), \, f=\left|
{{\begin{array}{*{20}c}
 f^{1} \hfill  \\
 f^{2} \hfill  \\
\end{array} }} \right|,$ was used. Such description [1--3] is adequate in the sense of its comparison with the nonrelativistic Schr$\mathrm{\ddot{o}}$dinger model of the electron. The direct relativistic analogue of the Schr$\mathrm{\ddot{o}}$dinger equation is the spinless Salpeter equation [4--6] for the one-component wave function: $i\partial_{t}f(x)=\sqrt{m^{2}-\Delta}f(x)$.

The relativistic equations mentioned above in an obvious way can be generalized for the particle multiplets with arbitrary spin. In the case of arbitrary spin and multi-component wave functions we suggested [7, 8] to call such type of equations as \textit{the Schr$\mathrm{\ddot{o}}$dinger--Foldy equations}. The important contribution of L. Foldy [1--3] into the consideration and analysis of the spinor field in the nonlocal canonical representation (and his analysis of the principles of heredity and correspondence with nonrelativistic quantum mechanics) was taken into account.

The analysis of the 4-component spinor field [1--3] enabled the authors of these papers to discern the quantum-mechanical interpretation of the Dirac equation. For these purposes the Dirac equation was transformed into another representation $i\partial_{t}f(x)=\gamma^{0}\sqrt{m^{2}-\Delta}f(x)$, which today is called the Foldy--Wouthuysen (FW) representation.  

In this article the Schr$\mathrm{\ddot{o}}$dinger--Foldy equation for the 4-component wave function (in all essential details) is under consideration. The axiomatic formulation of corresponding relativistic canonical quantum mechanics (RCQM) is presented briefly. The derivation of the Dirac equation directly from the Schr$\mathrm{\ddot{o}}$dinger--Foldy equation (without any additional assumptions) is given.           

We started the consideration of this subject in the paper [9].
Here the basic principles of the RCQM for the spin $s=\frac{1}{2}$ doublet and the derivation of
the Dirac equation from this model are under further
consideration. The foundations of RCQM were given in [1--4, 7--9]. Here the mathematically well-defined
consideration on the level of modern axiomatic approaches to the field theory [10] is provided.

In Sec. 2	a brief review of the different ways of the Dirac equation derivation is given. Our motivation and goals are considered.

In Sec. 3 the main notations and definitions are fixed.

In Sec. 4 the difference between the Schr$\mathrm{\ddot{o}}$dinger--Foldy equation and the FW equation is demonstrated. The advantages of the Schr$\mathrm{\ddot{o}}$dinger--Foldy equation in the detailed quantum-mechanical description of the fermionic doublet are demonstrated. The reasons of our postulation of the Schr$\mathrm{\ddot{o}}$dinger--Foldy equation to be the RCQM equation of motion are shown.

In Sec. 5 the brief axiomatic foundations of the RCQM are formulated.

In Sec. 6 we present our derivation of the Dirac equation. We start from the main principles of RCQM (directly from the Schr$\mathrm{\ddot{o}}$dinger--Foldy equation). 

In Sec. 7 we formulate the main conclusions and discuss briefly the negative point of view of RCQM for the possibility of negative mass of antiparticle.  

\section{Motivation and goals}

The significance of the Dirac equation and its wide-range
application in different models of theoretical physics (QED, QHD,
theoretical atomic and nuclear physics, solid systems) is
well-known. Let us recall only that the first analysis of this
equation enabled Dirac to give a theoretical prediction of the
positron, which was discovered experimentally by Anderson in
1932. The recent well-known application of the massless Dirac
equation to the graffen ribbons is an example of possibilities of
this equation. In our recent publications [11--15] we were able to
extend the domain of the Dirac equation application. We proved
[11--15] that this equation has not only fermionic but also the
bosonic features, can describe not only fermionic, but also the
bosonic states.

Therefore, the new ways of the derivation of the Dirac equation
are the actual problems. The new ways of the Dirac equation
derivation automatically visualize the ground principles, which
are in the foundations of the description of the elementary
particles on the basis of this equation. Hence, the active
consideration of the different ways of the Dirac equation
derivation is the subject of many contemporary publications. The
start from the different basic principles and assumptions is
considered.

Below a brief review of the different ways of the Dirac equation derivation is given.

 It is necessary to mark the elegant derivation given by Paul Dirac in his book [16]. Untill today it is very interesting for readers to feel the Dirac's thinking and to follow his logical steps. Nevertheless, the Dirac's consideration of the Schr$\mathrm{\ddot{o}}$dinger--Foldy equation, which was essentially used in his derivation [16], was not correct. Especially his assertion that Schr$\mathrm{\ddot{o}}$dinger--Foldy equation is unsatisfactory from the point of view of the relativistic theory. The Dirac's doubts were overcomed in [1--3]. Today it is confirmed by more then one hundred publications about FW and the spinless Salpeter equations, which have wide-range application in contemporary theoretical physics.  

 In well-known book [17] one can find excellent review of the Dirac theory and two different ways of the
Dirac equation derivation. At first, it is the presentation of the
Klein--Gordon equation in the form of the first-order
differential system of equations, the factorization of the
Klein--Gordon operator. Secondly, the Lagrange approach is considered and the Dirac equation is derived
from the variational Euler--Lagrange least action principle.

 In van der Waerden--Sakurai derivation [18] of the Dirac equation the spin of the electron is incorporated into the nonrelativistic theory. The representation of the nonrelativistic kinetic energy operator of the free spin 1/2 particle in the form $H^{\mathrm{KE}} = (\overrightarrow{\sigma} \cdot \overrightarrow{p})(\overrightarrow{\sigma} \cdot \overrightarrow{p})/2m$ and the relativistic expression $E^{2}-\overrightarrow{p}^{2}=m^{2}$ are used. After that the procedure of transition from 2-component to 4-component equation is fulfilled and explained.

 In the book [19] (second edition) the Dirac equation is
derived from the manifestly covariant transformational properties
of 4-component spinor.

 The derivation of the Dirac equation from the initial
geometric properties of the space-time and electron together with
wide-range discussion of the geometric principles of the electron
theory is the main content of the book [20]. The ideas of V. Fock
and D. Iwanenko [21, 22] on the geometrical sense of Dirac
$\gamma$-matrices are in the basis of the approach.

 It should be pointed out the derivation of the Dirac equation
based on the Bargman--Wigner classification of the irreducible unitary
representations of the Poincar$\mathrm{\acute{e}}$ group, see, e.
g. [23]. It is the illustrative demonstration of the possibilities
of the group-theoretical approach for the elementary particle
physics.

 In L. Foldy's papers [1--3] one can easy find the inverse problem, in which the Dirac equation is obtained from the FW equation. Nevertheless, it is only the transition from one representation of the spinor field to another.

 H. Sallhofer [24, 25] derived the Dirac equation for the hydrogen spectrum on the basis of start from the Maxwell equations in medium. Strictly speaking, only the stationary equations were considered.

 In the paper [26] quaternion measurable processes was introduced and the Dirac equation was derived from the Langevin equation associated with a two-valued process.

 The author of [27] was able to derive the Dirac equation from the conservation law of spin 1/2 current. The requirement that this current is conserved leads to a unique determination of the Lorentz invariant equation satisfied by the relativistic spin 1/2
field. Let us briefly comment that the complete list of conservation laws for the Dirac theory is the Noether consequence of the Dirac equation. Therefore, the validity of the inverse problem is really expected. Can it be considered as the independent derivation?  

 The Dirac equation has been derived [28] from the master equation of Poisson process by analytic continuation. The extension to
the case where a particle moves in an external field was given. It was shown, that the generalized master equation is intimately
connected with three-dimensional Dirac equation in an external field.

 In the paper [29] a method of deriving the Dirac equation from the relativistic Newton's second law was suggested. Such derivation is possible in a new formalism, which connects the special form of relativistic mechanics with the quantum mechanics. The author suggested a concept: velocity field. At first, the relativistic Newton’s second law was rewritten as a field equation in terms of the velocity field, which directly reveals a new relationship connecting to the quantum mechanics. After that it was shown that the Dirac equation can be derived from the field equation in a rigorous and consistent manner.

 In the paper [30] a geometrical derivation of the Dirac equation, by considering a spin 1/2 particle traveling
with the speed of light in a cubic spacetime lattice, was given. The mass of the particle acts to flip the
multi-component wavefunction at the lattice sites. Starting with a difference equation for the
case of one spatial and one time dimensions, the authors generalize the approach to higher dimensions.
Interactions with external electromagnetic and gravitational fields are also considered. Nevertheless, the idea of such derivation is based on the Dirac's observation that the instantaneous velocity operators of the spin 1/2 particle (hereafter called by the generic name “the electron”) have eigenvalues $\pm c$. This mistake of Dirac was demonstrated and overcomed in [1].

 Using the mathematical tool of Hamilton's bi-quaternions, the authors of [31] propose a derivation of the Dirac equation from a geodesic equation. Such derivation is given in the program of application of the theory of scale relativity to microphysics aims at recovering quantum mechanics as a new non-classical mechanics on a non-derivable space-time.

 M. Evans was successful to express his equation of general relativity (generally covariant field equation for gravitation and electromagnetism [32]) in spinor form, thus producing the Dirac equation in general relativity [33]. The Dirac equation in special relativity is recovered in the limit of Euclidean or flat spacetime.

 Ten years ago we already presented our own derivation of the Dirac equation [34--36]. The Dirac equation was derived from the slightly generalized Maxwell equations with gradient-like current and charge densities. This form of the Maxwell equations, which is directly linked with Dirac equation, is the maximally symmetrical variant of this equations. Such Maxwell equations are invariant with respect to 256-dimensional algebra (the well-known algebra of conformal group has only 15 generators). Of course, we derived only massless Dirac equation.

 Today we present new derivation of the Dirac equation. We derive the Dirac equation from the 4-component Schr$\mathrm{\ddot{o}}$dinger--Foldy equation of the RCQM. We postulate the Schr$\mathrm{\ddot{o}}$dinger--Foldy equation and construct the corresponding formalism of RCQM as the most fundamental model of fermionic doublet. At first, the brief axiomatic formulation of the RCQM foundations is given. After that the operator, which transforms the Schr$\mathrm{\ddot{o}}$dinger--Foldy equation into the Dirac equation, is given. Therefore, the new way of the Dirac equation derivation is presented.

Our main goal is following.  

To answer the question "Whether there exists a more fundamental model of a
"particle doublet" (as an elementary fundamental object), from which
the Dirac equation (and its content) would follow directly and unambiguously?"
We are able to demonstrate that axiomatically formulated RCQM of a
particle-antiparticle doublet of spin $s=\frac{1}{2}$ should be
chosen as such a model. In the text below the concrete and detailed illustration of this assertion on the
example of electron-positron doublet, $e^{-}e^{+}$-doublet, is
given.

\section{Notations and main definitions}

The model of RCQM for the elementary particle with $m>0$ and spin $s=\frac{1}{2}$, which satisfy the Schr$\mathrm{\ddot{o}}$dinger--Foldy equation $i\partial_{t}f(x)=\sqrt{m^{2}-\Delta}f(x); \, x\in \mathrm{M}(1,3), \, f=\left|
{{\begin{array}{*{20}c}
 f^{1} \hfill  \\
 f^{2} \hfill  \\
\end{array} }} \right|, \, \int d^{3}x\left|\varphi(x)\right|^{2}<\infty$, was suggested and approved in [1--3]. This model can be easily generalized to the case of arbitrary $\overrightarrow{s}$-multiplet, i. e. the "elementary object" with mass $m$ and spin $\overrightarrow{s} \equiv\left(s^{j}\right)=\left(s_{23},s_{31},s_{12}\right): \, \left[s^{j},s^{l}\right]=i\varepsilon^{jln}s^{n}$, where $\varepsilon^{jln}$ is the Levi-Civita tensor and $s^{j}=\varepsilon^{j\ell n}s_{\ell n}$ are the Hermitian $\mathrm{M}\times \mathrm{M}$ matrices -- the generators of M-dimensional representation of the spin group SU(2) (universal covering of the SO(3)$\subset$SO(1,3) group).

In this article we present the detalization of such generalization at the example of the spin $s=\frac{1}{2}$ Fermionic doublet. All mathematical and physical details of consideration, related to the choice of the concrete form of the spin $\overrightarrow{s}$ doublet, at the example of $e^{-}e^{+}$-doublet are illustrated.

We choose here the standard relativistic concepts, definitions and notations
in the convenient for our consideration form. For example, in the
Minkowski space-time

\begin{equation}
\label{eq1}
\mathrm{M}(1,3)=\{x\equiv(x^{\mu})=(x^{0}=t, \,
\overrightarrow{x}\equiv(x^{j}))\}; \quad \mu=\overline{0,3}, \,
j=1,2,3,
\end{equation}

\noindent the $x^{\mu}$ are the Cartesian (covariant)
coordinates of the points of the physical space-time in the
arbitrary-fixed inertial frame of references (IFR). We use the system of units $\hbar=c=1$. The metric tensor is given by

\begin{equation}
\label{eq2}
g^{\mu\nu}=g_{\mu\nu}=g^{\mu}_{\nu}, \, \left(g^{\mu}_{\nu}\right)=\mathrm{diag}\left(1,-1,-1,-1\right); \quad x_{\mu}=g_{\mu\nu}x^{\mu},
\end{equation}

\noindent the summation over the twice repeated index is implied.

The analysis of the relativistic invariance of an arbitrary physical model demands as a first step the consideration of its invariance with respect to the proper ortochronous Lorentz $\mbox{L}_ + ^ \uparrow $ = SO(1,3)=$\left\{\Lambda=\left(\Lambda^{\mu}_{\nu}\right)\right\}$ and  Poincar$\mathrm{\acute{e}}$ $\mbox{P}_ + ^
\uparrow = \mbox{T(4)}\times )\mbox{L}_ + ^ \uparrow  \supset \mbox{L}_ + ^ \uparrow$ groups. This invariance in an arbitrary relativistic model is the realization of the Einstein's relativity principle in the form of special relativity. Note that the mathematical correctness demands to consider the invariance mentioned above as the invariance with respect to the universal coverings $\mathcal{L}$ = SL(2,C) and $\mathcal{P}\supset\mathcal{L}$ of the groups $\mbox{L}_ + ^ \uparrow $ and $\mbox{P}_ + ^ \uparrow $, respectively.

For the group $\mathcal{P}$ we choose the real parameters $a=\left(a^{\mu}\right)\in$M(1,3) and $\varpi\equiv\left(\varpi^{\mu\nu}=-\varpi^{\nu\mu}\right)$, which physical meaning is well-known. For the standard $\mathcal{P}$ generators $\left(p_{\mu},j_{\mu\nu}\right)$ we use the commutation relations in the manifestly covariant form

\begin{equation}
\label{eq3}
\left[p_{\mu},p_{\nu}\right]=0, \, \left[p_{\mu},j_{\rho\sigma}\right]=ig_{\mu\rho}p_{\sigma}-ig_{\mu\sigma}p_{\rho}, \, \left[j_{\mu\nu},j_{\rho\sigma}\right]=-i\left(g_{\mu\rho}j_{\nu\sigma}+g_{\rho\nu}j_{\sigma\mu}+g_{\nu\sigma}j_{\mu\rho}+g_{\sigma\mu}j_{\rho\nu}\right).
\end{equation}

\section{Canonical equation of motion of relativistic quantum mechanics}

In this section we consider the comparison of Schr$\mathrm{\ddot{o}}$dinger--Foldy and FW equations for the fermionic doublet. On this bases we demonstrate why the Schr$\mathrm{\ddot{o}}$dinger--Foldy equation should be chosen as the main equation of motion in RCQM.

The Schr$\mathrm{\ddot{o}}$dinger--Foldy equation for the fermionic spin 1/2 doublet is given by

\begin{equation}
\label{eq4}
i\partial_{t}f(x)=\sqrt{m^{2}-\Delta}f(x), 
\end{equation}

\noindent where

\begin{equation}
\label{eq5}
f\equiv \mathrm{column}(f^{1},f^{2},f^{3},f^{4}).
\end{equation}

\noindent This equation, similarly to the nonrelativistic 4-component Schr$\mathrm{\ddot{o}}$dinger equation

\begin{equation}
\label{eq6}
i\partial_{t}f(x)=\frac{\overrightarrow{p}^{2}}{2m}f(x)
\end{equation}

\noindent (involving also the internal degrees of freedom, spin etc.) is considered in the quantum-mechanical Hilbert space

\begin{equation}
\label{eq7}
\mathrm{H}^{3,4}=\mathrm{L}_{2}(\mathrm{R}^3)\otimes\mathrm{C}^{\otimes4}=\{f=(f^{\alpha}):\mathrm{R}^{3}\rightarrow\mathrm{C}^{\otimes4};
 \, \int
d^{3}x|f(t,\overrightarrow{x})|^{2} <\infty\},
\end{equation}

\noindent where $d^{3}x$ is the Lebesgue measure in the space $\mathrm{R}^{3}\subset \mathrm{M}(1,3)$ of the eigenvalues of the position operator $\overrightarrow{x}$ of the Cartesian coordinate of the doublet in an arbitrary-fixed inertial frame of reference (IFR). In (4)--(7) and below the two upper components $f^{1}, \, f^{2}$ of the vector $f\in\mathrm{H}^{3,4}$ are the components of the electron wave function $\varphi_{-}$ and the two lower components $f^{3}, \, f^{4}$ are those of the positron wave function $\varphi_{+}$.

The general solution of the Schr$\mathrm{\ddot{o}}$dinger--Foldy equation (4), similarly to the general solution of the nonrelativistic 4-component Schr$\mathrm{\ddot{o}}$dinger equation (6), is given by

$$f(x)= \left|
{{\begin{array}{*{20}c}
 f_{e^{-}} \hfill  \\
 f_{e^{+}} \hfill  \\
\end{array} }} \right| =$$

\begin{equation}
\label{eq8}
\frac{1}{\left(2\pi\right)^{\frac{3}{2}}}\int d^{3}k e^{-ikx}\left[a^{-}_{+}(\overrightarrow{k})\mathrm{d}_{1}+a^{-}_{-}(\overrightarrow{k})\mathrm{d}_{2}+a^{+}_{-}(\overrightarrow{k})\mathrm{d}_{3}+a^{+}_{+}(\overrightarrow{k})\mathrm{d}_{4}\right],
\end{equation}

\noindent where 

\begin{equation}
\label{eq9}
kx\equiv \omega t -\overrightarrow{k}\overrightarrow{x}, \quad \omega \equiv \sqrt{\overrightarrow{k}^{2}+m^{2}},
\end{equation}

\noindent the 4-columns $\mathrm{d}_{\alpha}$ are the Cartesian orts in the space
$\mathrm{C}^{\otimes4}\subset\mathrm{H}^{3,4}$

\begin{equation}
\label{eq10}
\mathrm{d}_{1} = \left|
\begin{array}{cccc}
 1 \\
 0 \\
0 \\
0 \\
\end{array} \right|, \, \mathrm{d}_{2} = \left|
\begin{array}{cccc}
 0 \\
 1 \\
0 \\
0 \\
\end{array} \right|, \, \mathrm{d}_{3} = \left|
\begin{array}{cccc}
 0 \\
 0 \\
1 \\
0 \\
\end{array} \right|, \,\mathrm{d}_{4} = \left|
\begin{array}{cccc}
 0 \\
 0 \\
0 \\
1 \\
\end{array} \right|, 
\end{equation}

\noindent the functions $a^{-}_{+}(\overrightarrow{k}),\,a^{-}_{-}(\overrightarrow{k})$
are the quantum-mechanical momentum-spin amplitudes of the
particle with charge -e and eigen values of spin projection +1/2
and -1/2;
$a^{+}_{-}(\overrightarrow{k}),\,a^{+}_{+}(\overrightarrow{k})$
are the quantum-mechanical momentum-spin amplitudes of the
antiparticle with charge +e and eigen values of spin projection
-1/2 and +1/2, respectively.

In the general solution (8) we use the modern experimentally verified understanding of the positron as the "mirror mapping" of the electron. Such understanding leads to the specific postulation of the explicit forms of the charge sign and spin operators (see the fprmulae (24) in [7]), which determine the form of the solution (8).

Contrary to the Schr$\mathrm{\ddot{o}}$dinger--Foldy equation (4), the FW equation has the form

\begin{equation}
\label{eq11}
i\partial_{t} \phi(x)=\gamma^{0}\sqrt{m^{2}-\Delta} \phi(x), \quad \gamma^{0}= \left| {{\begin{array}{*{20}c}
 \mathrm{I}_{2} \hfill & 0 \hfill \\
 0 \hfill & -\mathrm{I}_{2} \hfill \\
\end{array} }} \right|, \quad \mathrm{I}_{2}=\left|
{{\begin{array}{*{20}c}
 1 \hfill & 0 \hfill \\
 0 \hfill & { 1} \hfill \\
\end{array} }} \right|,
\end{equation}

\noindent and its general solution is given by

$$\phi(x)=\left|
{{\begin{array}{*{20}c}
 f_{e^{-}} \hfill  \\
f^{*}_{e^{+}} \hfill  \\
\end{array} }} \right|=\frac{1}{(2\pi)^{\frac{3}{2}}}\int
d^{3}k\{e^{-ikx}[a^{-}_{+}(\overrightarrow{k})\mathrm{d}_{1}+a^{-}_{-}(\overrightarrow{k})\mathrm{d}_{2}]$$
\begin{equation}
\label{eq12}
+e^{ikx}[a^{*+}_{-}(\overrightarrow{k})\mathrm{d}_{3}+a^{*+}_{+}(\overrightarrow{k})\mathrm{d}_{4}]\}.
\end{equation}

The equations (4) and (11) are different due to the presence of $\gamma^{0}$ in (11). Owing to this fact the general solutions (8) and (12) are also different. The solution (8) is the direct sum of the electron $f_{e^{-}}$ and the positron $f_{e^{+}}$ quantum-mechanical wave functions. The  solution (12) is the direct sum of the electron $f_{e^{-}}$ and the complex conjugated positron $f^{*}_{e^{+}}$ wave functions. Furthermore, the solution (8) contains only positive energetic states both for the electron and positron, whereas the solution (12) contains the positive energetic states of the electron and negative energetic states of the positron.

Note that solution (8) for the wave function of spin 1/2 doublet is expressed in the terms of relativistic de Broglie waves for the electron and the positron 

\begin{equation}
\label{eq13}
\varphi_{\vec{k}\mathrm{A}}(t,\overrightarrow{x})=\frac{1}{(2\pi)^{\frac{3}{2}}}e^{-i\omega t + i\vec{k}\vec{x} }\mathrm{d}_{\mathrm{A}}, \quad  \mathrm{A} = 1,2,3,4.
\end{equation}

\noindent The expressions (13) are the fundamental (basis) solutions of the equation (4), they do not belong to the Hilbert space (7) (their $\mathrm{H}^{3,4}$-norms are equal to the infinity). The mathematical correctness of the consideration is ensured by the applying of the rigged Hilbert space

\begin{equation}
\label{eq14}
\mathrm{S}^{3,4}\equiv \mathrm{S}(\mathrm{R}^{3})\times\mathrm{C}^{4}\subset\mathrm{H}^{3,4}\subset\mathrm{S}^{3,4*}.
\end{equation}

\noindent Here $\mathrm{S}^{3,4}$ is the 4-component Schwartz test function space over the space $\mathrm{R}^{3}\subset \mathrm{M}(1,3)$, and $\mathrm{S}^{3,4*}$ is the space of 4-component Schwartz generalized functions, which is conjugated to the Schwartz test function space $\mathrm{S}^{3,4}$
by the corresponding topology (see, e. g., [38]). Strictly speaking, the mathematical correctness of consideration demands to make the calculations in the space $\mathrm{S}^{3,4*}$ of generalized functions, i. e. with the application of cumbersome functional analysis.

Nevertheless, let us take into account that the Schwartz test function space $\mathrm{S}^{3,4}$ in the
triple (22) is \textit{kernel}. It means that $\mathrm{S}^{3,4}$ is dense both in quantum-mechanical space $\mathrm{H}^{3,4}$ and in the space of generalized functions $\mathrm{S}^{3,4*}$. Therefore, any physical state $f\in\mathrm{H}^{3,4}$ can be approximated with an arbitrary precision by the corresponding elements of the Cauchy sequence in $\mathrm{S}^{3,4}$, which converges to the given $f\in\mathrm{H}^{3,4}$ or to the given $f\in\mathrm{S}^{3,4*}$. Further, taking into account the requirement to measure the arbitrary value of the model with non-absolute precision, it means that all concrete calculations can be fulfilled within the Schwartz test function space $\mathrm{S}^{3,4}$. 

Note that if the general solution $f(x)$ (8) belongs to the $\mathrm{S}^{3,4}$ then the amplitudes $a^{-}_{+}(\overrightarrow{k}),\,a^{-}_{-}(\overrightarrow{k}), \, a^{+}_{-}(\overrightarrow{k}), \, a^{+}_{+}(\overrightarrow{k})$ also belong to the test function space $\mathrm{S}(\mathrm{R}^{3}_{\vec{k}})$. It is enough for the approximation of any experimental situation.

Contrary to this situation, even if amplitudes ${a}$ in (12) belong to the $\mathrm{S}(\mathrm{R}^{3}_{\vec{k}})$, the general solution (12) of the FW equation (11) does not belong to the quantum-mechanical Hilbert space (7) (due to the indefinite metric in the space of solutions (12)). Nevertheless, the mathematical correctness of consideration is achieved (ensured) in the space $\mathrm{S}^{3,4}\subset \mathrm{S}^{3,4*}$ due to the fact that $\mathrm{S}^{3,4}$ is dense in $\mathrm{S}^{3,4*}$.

Therefore, we are able to demonstrate the difference between the Schr$\mathrm{\ddot{o}}$dinger--Foldy equation (4) and the FW equation (11) in the quantum-mechanical description of the fermionic doublet. In despite of the fact that in the FW representation one can discern some quantum-mechanical aspects, however, in general the FW equation does not guarantee the detailed quantum-mechanical description of the fermionic doublet (as well as the Dirac equation).

Hence, we postulate the Schr$\mathrm{\ddot{o}}$dinger--Foldy equation (4) to be the RCQM equation of motion for the fermionic spin 1/2doublet and construct the corresponding canonical quantum-mechanical formalism. 

\section{Relativistic canonical quantum mechanics of the Fermi-doublet}

The \textit{axioms of the model} are formulated on the level of correctness of von Neuman's monograph [37]. Requirements of such physically verified principles as \textit{the principle of relativity with respect to the tools of cognition (PRTC)}, \textit{principle of heredity (PH)} with both classical mechanics of single mass point and non-relativistic quantum mechanics (and \textit{the principle of correspondence (PС)} with these theories), and also \textit{the Einstein principle of relativity (EPR)}, are taken into consideration. The last principle requires first of all \textit{the special relativity (SR)} to be taken into account.

The \textit{ basic axioms} of the model (we present here the brief consideration) as the mathematical assertions have the form.

\textbf{On the space of states}. The space of states of isolated  $e^{-}e^{+}$-doublet in arbitrarily-fixed inertial frame of reference (IFR) in its  $\overrightarrow{x}$-realization is the Hilbert space $\mathrm{H}^{3,4}$ (7) of complex-valued 4-component square-integrable functions of $x\in\mathrm{R}^{3}\subset \mathrm{M}(1,3)$ (similarly, in momentum,  $\overrightarrow{p}$-realization). Here  $\overrightarrow{x}$ and $\overrightarrow{p}$ are the operators of canonically conjugated dynamical variables of $e^{-}e^{+}$-doublet, and the vectors $f$, $\tilde{f}$  in $\overrightarrow{x}$ and $\overrightarrow{p}$ realizations are linked by the 3-dimensional Fourier transformation (the variable $t$ is the parameter of time-evolution).

\textbf{The mathematical correctness of the consideration} demands the application of the rigged Hilbert space (14), where the Schwartz test function space $\mathrm{S}^{3,4}$, which is the verified tool of the PRTC realization, is kernel (i. e., it is dense both in $\mathrm{H}^{3,4}$ and in the space $\mathrm{S}^{3,4*}$ of the generalized Schwartz functions). Such application allows us to fulfill, without any loss of generality, all necessary calculations in the space $\mathrm{S}^{3,4}$ on the level of correct differential and integral calculus. The more detailed consideration is given in paragraphs after the formula (14).

\textbf{On the time evolution of the state vectors}. The time dependence of the state vectors $f\in \mathrm{H}^{3,4}$  (time t is the parameter of evolution) is given either in the integral form by the unitary operator

\begin{equation}
\label{eq15}
u\left(t_{0},t\right)=\exp \left[-i \widehat{\omega}(t-t_{0})\right]; \quad \widehat{\omega}\equiv\sqrt{-\Delta +m^{2}},
\end{equation}

\noindent (below is put $t_{0}=t$), or in the differential form by the Schr$\mathrm{\ddot{o}}$dinger--Foldy equation of motion (4). Here the operator $\widehat{\omega}\equiv\sqrt{-\Delta +m^{2}}$  is the relativistic analog of the energy operator (Hamiltonian) of nonrelativistic quantum mechanics. The Minkowski space-time M(1,3) is pseudo Euclidean with metric $g =\mathrm{diag}(+1,-1,-1,-1)$.

\textbf{On the fundamental dynamical variables}. The dynamical variable $\overrightarrow{x}\in \mathrm{R}^{3}\subset$M(1,3)(as well as the variable $\overrightarrow{k}\in \mathrm{R}_{\vec{k}}^{3}$) represents the external degrees of freedom of $e^{-}e^{+}$-doublet. The spin $\overrightarrow{s}$ of $e^{-}e^{+}$-doublet is the first in the list of the carriers of the internal degrees of freedom. Taking into account the Pauli principle and the fact, that experimentally positron is observed as the mirror reflection of the electron, the operators of  the charge sign  and the spin of $e^{-}e^{+}$-doublet are taken in the form

\begin{equation}
\label{eq16} g\equiv-\gamma^0 = \left| {{\begin{array}{*{20}c}
 -\mathrm{I}_{2} \hfill & 0 \hfill \\
 0 \hfill & \mathrm{I}_{2} \hfill \\
\end{array} }} \right|, \quad \overrightarrow{s} =
\frac{1}{2}\left| {{\begin{array}{*{20}c}
 \overrightarrow{\sigma} \hfill  0 \hfill \\
 0 -C\hfill\overrightarrow{\sigma}\hfill C \\
\end{array} }} \right|, \quad \mathrm{I}_{2}=\left|
{{\begin{array}{*{20}c}
 1 \hfill & 0 \hfill \\
 0 \hfill & { 1} \hfill \\
\end{array} }} \right|,
\end{equation}

\noindent where $\overrightarrow{\sigma}$ are the standard Pauli matrices and $C$ is the operator of complex conjugation. The spin matrices (16) satisfy the commutation relations of the algebra of SU(2) group

\begin{equation}
\label{eq17}
\overrightarrow{s} \equiv\left(s^{j}\right)=\left(s_{23},s_{31},s_{12}\right): \, \left[s^{j},s^{l}\right]=i\varepsilon^{jln}s^{n}; \, \varepsilon^{123}=+1.
\end{equation}

\noindent where $\varepsilon^{jln}$ is the Levi-Civita tensor and $s^{j}=\varepsilon^{j\ell n}s_{\ell n}$ are the Hermitian $\mathrm{4}\times \mathrm{4}$ matrices -- the generators of 4-dimensional reducible representation of the spin group SU(2) (universal covering of the SO(3)$\subset$SO(1,3) group).

\textbf{On the algebra of observables}. Using the operators of canonically conjugated coordinate $\overrightarrow{x}$ and momentum $\overrightarrow{p}$ (where $\left[x^{j}, p^{\ell}\right]=i\delta_{j\ell} $ in $\mathrm{H}^{3,4}$), being completed by the operators $\overrightarrow{s}$ and $g$, we construct the algebra of observables (according to the PH) as the Hermitian functions of 10 ($\overrightarrow{x}, \, \overrightarrow{p} \, \overrightarrow{s}, \, -\gamma^{0}$) generating elements of the algebra.

\textbf{On the relativistic invariance of the theory}. This invariance (realization of the SR) is ensured by the proof of the invariance of the Schr$\mathrm{\ddot{o}}$dinger--Foldy equation (4) with respect to the unitary representation of the universal covering $\mathcal{P}\supset\mathcal{L}$=SL(2,C) of the proper ortochronous Poincar$\mathrm{\acute{e}}$ group $\mbox{P}_ + ^
\uparrow = \mbox{T(4)}\times )\mbox{L}_ + ^ \uparrow  \supset \mbox{L}_ + ^ \uparrow$. Here $\mathcal{L}$ = SL(2,C) is the universal covering of proper ortochronous Lorentz group $\mbox{L}_ + ^ \uparrow $.

The generators of the fermionic $\mathcal{P}^{\mathrm{F}}$ representation of the group $\mathcal{P}$, with respect to which the Schr$\mathrm{\ddot{o}}$dinger--Foldy equation (4) is invariant, are given by

\begin{equation}
\label{eq18}
\widehat{p}_{0}=\widehat{\omega}\equiv \sqrt{-\Delta+m^{2}}, \, \widehat{p}_{\ell}=i\partial_{\ell}, \, \widehat{j}_{\ell n}=x_{\ell}\widehat{p}_{n}-x_{n}\widehat{p}_{\ell}+s_{ln}\equiv \widehat{m}_{\ell n}+s_{\ell n},
\end{equation}

\begin{equation}
\label{eq19}
\widehat{j}_{0 \ell}=-\widehat{j}_{\ell\ell 0}=t\widehat{p}_{\ell}-\frac{1}{2}\left\{x_{\ell},\widehat{\omega}\right\}-\left(\frac{s_{\ell n}\widehat{p}_{n}}{\widehat{\omega}+m} \equiv \breve{s}_{\ell}\right),
\end{equation}

\noindent in the $\overrightarrow{x}$-realization of the space $\mathrm{H}^{3,4}$ (7) and by

\begin{equation}
\label{eq20}
p_{0}=\omega, \, p_{\ell}=k_{\ell}, \, \widetilde{j}_{\ell n}=\widetilde{x}_{\ell}k_{n}-\widetilde{x}_{n}k_{\ell}+s_{\ell n}; \quad (\widetilde{x}_{\ell}=-i\widetilde{\partial}_{\ell}, \, \widetilde{\partial}_{\ell}\equiv \frac{\partial}{\partial k^{\ell}}),
\end{equation}

\begin{equation}
\label{eq21}
\widetilde{j}_{0 \ell}=-\widetilde{j}_{\ell 0}=t k_{l}-\frac{1}{2}\left\{\widetilde{x}_{\ell},\omega \right\}-\left(\frac{s_{\ell n}k_{n}}{\omega+m} \equiv \breve{\widetilde{s}}_{\ell}\right),
\end{equation}

\noindent in the momentum $\overrightarrow{k}$-realization $\widetilde{\mathrm{H}}^{3,4}$ of the doublet states space, respectively. The explicit form of the spin operators $s_{ln}$ in the formulae (18)--(21), which is used for the $e^{-}e^{+}$-doublet, is given in the formula (16).

In despite of manifestly non-covariant forms (18)--(21) of the $\mathcal{P}^{\mathrm{F}}$-generators, they satisfy the commutation relations of the $\mathcal{P}$ algebra in manifestly covariant form (3).

The $\mathcal{P}^{\mathrm{F}}$-representation of the group $\mathcal{P}$ in the space $\mathrm{H}^{3,4}$ (7) is given by the converged in this space exponential series

\begin{equation}
\label{eq22}
\mathcal{P}^{\mathrm{F}}: \, (a,\varpi)\rightarrow U(a,\varpi)=\exp (-ia^{0}\widehat{\omega}-i\overrightarrow{a}\widehat{\overrightarrow{p}}-\frac{i}{2}\varpi^{\mu\nu}\widehat{j}_{\mu\nu}),
\end{equation}

\noindent or, in the momentum space $\widetilde{\mathrm{H}}^{3,4}$, by corresponding exponential series given in terms of the generators (20), (21).

We emphasize that the modern definition of $\mathcal{P}$ invariance (or $\mathcal{P}$ symmetry) of the equation of motion (4) in $\mathrm{H}^{3,4}$ is given by the following assertion, see, e. g. [39]. \textit{The set} $\mathrm{F}\equiv\left\{f\right\}$ \textit{of all possible solutions of the equation (4) is invariant with respect to the} $\mathcal{P}^{\mathrm{F}}$-\textit{representation of the group} $\mathcal{P}$, \textit{if for arbitrary solution} $f$ \textit{and arbitrarily-fixed parameters} $(a,\varpi)$ \textit{the assertion}

\begin{equation}
\label{eq23}
(a,\varpi)\rightarrow U(a,\varpi)\left\{f\right\}=\left\{f\right\}\equiv\mathrm{F}
\end{equation}

\noindent \textit{is valid}. Furthermore, the assertion (23) is ensured by the fact that (as it is easy to verify) all the $\mathcal{P}$-generators (18), (19) commute with the operator $i\partial_{t}-\sqrt { - \Delta + m^2}$ of the equation (4). 

Not a matter of fact that in RCQM many manifestly noncovariant objects are used, \textit{the model under consideration is relativistic invariant in the sense of the definition given above}.

\textbf{On the main and additional conservation laws}.

Similarly to the nonrelativistic quantum mechanics \textit{the conservation laws are found in the form of quantum-mechanical mean values of the operators, which commute with the operator of the equation of motion}.

The important physical consequence of the assertion about the relativistic invariance is the fact that 10 integral dynamical variables of the doublet

\begin{equation}
\label{eq24}
(P_{\mu}, \, J_{\mu\nu}) \equiv \int d^{3}x f^{\dag}(t,\overrightarrow{x})(\widehat{p}_{\mu}, \, \widehat{j}_{\mu\nu})f(t,\overrightarrow{x})=\mathrm{Const}
\end{equation}

\noindent do not depend on time, i. e. they are the constants of motion for this doublet.

Note that the external and internal degrees of freedom  for the free $e^{-}e^{+}$-doublet are independent. Therefore, the operator $\overrightarrow{s}$ (16) commutes not only with the operators $\widehat{\overrightarrow{p}}, \overrightarrow{x}$, but also with the orbital part $\widehat{m}_{\mu\nu}$ of the total angular momentum operator. And both operators $\overrightarrow{s}$ and $\widehat{m}_{\mu\nu}$ commute with the operator $i\partial_{t}-\sqrt { - \Delta + m^2}$ of the equation (4). Therefore, besides the 10 main (consequences of the 10 Poincar$\mathrm{\acute{e}}$ generators) conservation laws (24), the 12 additional constants of motion exist for the free $e^{-}e^{+}$-doublet. These additional conservation laws are the consequences of the operators of the following observables:

\begin{equation}
\label{eq25}
s_{j}, \, \breve{s}_{\ell}=\frac{s_{\ell n}\widehat{p}_{n}}{\widehat{\omega}+m}, \, \widehat{m}_{\ell n}=x_{l}\widehat{p}_{n}-x_{n}\widehat{p}_{\ell}, \, \widehat{m}_{0 \ell}=-\widehat{m}_{\ell 0}=t\widehat{p}_{\ell}-\frac{1}{2}\left\{x_{\ell},\widehat{\omega}\right\}.
\end{equation}

\noindent where $s_{j}=s_{\ell n}$ are given in (16).

Thus, the following assertions can be proved. In the space $\mathrm{H}^{\mathrm{A}}=\left\{A\right\}$ of the quantum-mechanical amplitudes the 10 main conservation laws (24) have the form

\begin{equation}
\label{eq26}
(P_{\mu},J_{\mu\nu})=\int d^{3}k A^{\dag}(\overrightarrow{k})(\widetilde{p}_{\mu},\widetilde{j}_{\mu\nu})A(\overrightarrow{k}), \quad A(\overrightarrow{k})\equiv \left|
{{\begin{array}{*{20}c}
 a^{-}_{\mathrm{r}}\\
 a^{+}_{\acute{\mathrm{r}}}\\
\end{array} }} \right|,
\end{equation}

\noindent where the density generators of $\mathcal{P}^{\mathrm{A}}$, $(\widetilde{p}_{\mu},\widetilde{j}_{\mu\nu})$ of (26) are given by

\begin{equation}
\label{eq27}
\widetilde{p}_{0}=\omega, \, \widetilde{p}_{l}=k_{l}, \, \widetilde{j}_{ln}=\widetilde{x}_{l}k_{n}-\widetilde{x}_{n}k_{l}+s_{ln};\quad (\widetilde{x}_{l}=-i\widetilde{\partial}_{l}, \, \widetilde{\partial}_{l}\equiv \frac{\partial}{\partial k^{l}}),
\end{equation}

\begin{equation}
\label{eq28}
\widetilde{j}_{0l}=-\widetilde{j}_{l0}=-\frac{1}{2}\left\{\widetilde{x}_{l},\omega \right\}-(\breve{\widetilde{s}}_{l}\equiv\frac{s_{ln}k_{n}}{\omega+m}).
\end{equation}

\noindent In the formula (26) $A(\overrightarrow{k})\equiv \left|
{{\begin{array}{*{20}c}
 a^{-}_{\mathrm{r}}\\
 a^{+}_{\acute{\mathrm{r}}}\\
\end{array} }} \right|$ is the column of amplitudes $a^{-}_{+}(\overrightarrow{k}),\,a^{-}_{-}(\overrightarrow{k}), \, a^{+}_{-}(\overrightarrow{k}), \, a^{+}_{+}(\overrightarrow{k})$, where $\mathrm{r}=(+,-), \acute{\mathrm{r}}=(-,+)$.

Note that the operators (26)--(28) satisfy the Poincar$\mathrm{\acute{e}}$ commutation relations in the manifestly covariant form (3). 

It is evident that 12 additional conservation laws 

\begin{equation}
\label{eq29}
(M_{\mu \nu}, \, S_{\ell n}, \, \breve{S}_{\ell}) \equiv \int d^{3}x f^{\dag}(t,\overrightarrow{x})(\widehat{m}_{\mu \nu}, \, s_{\ell n}, \, \breve{s}_{\ell})f(t,\overrightarrow{x})=\mathrm{Const}
\end{equation}

\noindent generated by the operators (25), are the separate terms in the expressions (26)--(28) of principal (main) conservation laws.

\textbf{On the Clifford--Dirac algebra}.

The Clifford--Dirac algebra of the $\gamma$-matrices must be introduced into the FW representations. The reasons are as follows.

The part of the Clifford--Dirac algebra operators are directly related to the spin 1/2 doublet operators $(\frac{1}{2}\gamma^{2}\gamma^{3}, \,\frac{1}{2}\gamma^{3}\gamma^{1}, \,\frac{1}{2}\gamma^{1}\gamma^{2})$. In the FW representation for the spinor field these spin operators commute with the Hamiltonian and with the operator of the equation of motion (11). In the Pauli--Dirac representation these operators do not commute with the Dirac equation operator. Only the sums of the orbital and such spin operators commute with the Diracian. So \textit{if we want to relate the orts of the Clifford--Dirac algebra with the true spin we must introduce this algebra into the FW representation}.

In the quantum-mechanical representation (i. e. in the space of the solutions (8) of the Schr$\mathrm{\ddot{o}}$dinger--Foldy equation(4)) the $\gamma$-matrices are obtained by the transformation $v$ given below in the formulas (31) and in (33), (34) of Sec. 6. 

Moreover, we use the generalized Clifford--Dirac algebra over the field of real numbers. This algebra was introduced in the papers [11--15]. The use of 29 orts of this \textit{proper extended real Clifford--Dirac algebra} gives the additional possibilities in comparison with only 16 elements of the standard Clifford--Dirac algebra, see, e. g., [11--15].

The definitions of spin matrices (16) de facto determines so-called "quantum-mechanical" representation of the Dirac matrices

\begin{equation}
\label{eq30}
\bar{\gamma}^{\mu}: \, \bar{\gamma}^{\mu}\bar{\gamma}^{\nu}+\bar{\gamma}^{\nu}\bar{\gamma}^{\mu}=2g^{\mu\nu}; \, \bar{\gamma}^{-1}_{0}=\bar{\gamma}_{0}, \, \bar{\gamma}^{-1}_{l}=-\bar{\gamma}_{l},
\end{equation}

\noindent The matrices $\bar{\gamma}^{\mu}$ (30) of this representation are linked to the Dirac matrices $\gamma^{\mu}$ in the standard Pauli-Dirac (PD) representation:

\begin{equation}
\label{eq31}
\bar{\gamma}^{0}=\gamma^{0}, \, \bar{\gamma}^{1}=\gamma^{1}C, \, \bar{\gamma}^{2}=\gamma^{0}\gamma^{2}C, \, \bar{\gamma}^{3}=\gamma^{3}C, \, \bar{\gamma}^{4}=\gamma^{0}\gamma^{4}C; \quad \bar{\gamma}^{\mu}=v\gamma^{\mu}v, \quad v \equiv\left|
{{\begin{array}{*{20}c}
 \mathrm{I}_{2} \hfill & 0 \hfill \\
 0 \hfill &  C\mathrm{I}_{2} \hfill \\
\end{array} }} \right|=v^{-1},
\end{equation}

\noindent where the standard Dirac matrices $\gamma^{\mu}$ are given by

\begin{equation}
\label{eq32}
\gamma^0 = \left| {{\begin{array}{*{20}c}
 \mathrm{I}_{2} \hfill & 0 \hfill \\
 0 \hfill & -\mathrm{I}_{2} \hfill \\
\end{array} }} \right|, \quad \gamma ^k = \left| {{\begin{array}{*{20}c}
 0 \hfill & {\sigma ^k} \hfill \\
 { - \sigma ^k} \hfill & 0 \hfill \\
\end{array} }} \right|, \quad \mu=0,1,2,3.
\end{equation}

\noindent Note that in the terms of $\bar{\gamma}^{\mu}$ matrices (31) the spin operator (16) have the form $\overrightarrow{s}=\frac{i}{2}(\bar{\gamma}^{2}\bar{\gamma}^{3}, \, \bar{\gamma}^{3}\bar{\gamma}^{1}, \, \bar{\gamma}^{1}\bar{\gamma}^{2})$.

The $\bar{\gamma}^{\mu}$ matrices (31) together with the matrix $\bar{\gamma}^{4}\equiv \bar{\gamma}^{0}\bar{\gamma}^{1}\bar{\gamma}^{2}\bar{\gamma}^{3}$, imaginary unit $i\equiv\sqrt{-1}$ and operator $C$ of complex conjugation in $\mathrm{H}^{3,4}$ generate the quantum-mechanical representations of the extended real Clifford-Dirac algebra and proper extended real Clifford-Dirac algebra, which were put into consideration in [11] (see also [12--15]).

\textbf{On the principles of the heredity and the correspondence}. The explicit forms (24)-(29) of the main and additional conservation laws demonstrate evidently that the model of RCQM satisfies the principles of the heredity and the correspondence with the non-relativistic classical and quantum theories. The deep analogy between RCQM and these theories for the physical system with the finite number degrees of freedom (where the values of the free dynamical conserved quantities are additive) is also evident.

\section{Derivation of the Foldy-Wouthuysen and the standard Dirac equations}

We consider briefly the derivation of the FW and the Dirac equations on the basis of the start from the the Schr$\mathrm{\ddot{o}}$dinger--Foldy equation (4). That means the Dirac equation is the consequence of the quantum-mechanical spin 1/2 doublet model.

The link between the the Schr$\mathrm{\ddot{o}}$dinger--Foldy equation equation (4) and the FW equation (11) is given by the operator $v$

\begin{equation}
\label{eq33}
 v=\left|
{{\begin{array}{*{20}c}
 \mathrm{I}_{2} \hfill & 0 \hfill \\
 0 \hfill &  C\mathrm{I}_{2} \hfill \\
\end{array} }} \right|; \quad v^{2}=\mathrm{I}_{4}, \quad \mathrm{I}_{2}=\left|
{{\begin{array}{*{20}c}
 1 \hfill & 0 \hfill \\
 0 \hfill & { 1} \hfill \\
\end{array} }} \right|,
\end{equation}

\noindent  $C$ is the operator of complex conjugation, the operator of involution in the space $\mathrm{H}^{3,4}$. The operator $v$ (33) transforms arbitrary operator $q$ of the RCQM into the operator $Q$ in the FW representation for the spinor field \textit{and vice versa}: 

\begin{equation}
\label{eq34}
 Q=vqv\leftrightarrow q=vQv.
\end{equation}

\noindent The only warning is that formula (34) is valid only for the anti-Hermitian operators! It means that in order to avoid the mistakes one must apply this formula only for the prime (anti-Hermitian) energy-momentum, angular momentum and spin quantities. Justification and use of the conception of the prime generators of the Lie groups are given in [11--15].

The role of anti-Hermitian operators in physics is well-known. As well as the physical parameters of groups and algebras are real, then it is convenient to associate with them just the anti-Hermitian generators. For example, the real parameters $a^{\mu}, \, \varpi^{\mu\nu}$ of translations and rotations of the Poincar$\mathrm{\acute{e}}$ group are associated with the anti-Hermitian generators $\widehat{p}_{\mu}, \, \widehat{j}_{\mu\nu}$, where $\widehat{p}_{\mu}=\partial_{\mu}$, etc. The mathematical correctness of appealing to the anti-Hermitian generators is considered in details in [40, 41]. In our papers just the use of the anti-Hermitian generators allowed us [11--15] to find the additional bosonic properties of the FW and Dirac equations. The details are not the subject of this consideration.

Here, in order to work with mathematically well-defined relationship between the Schr$\mathrm{\ddot{o}}$dinger--Foldy and FW equation we slightly rewrite these equations and present them in completely equivalent forms in the terms of the anti-Hermitian operators. Thus, we consider the Schr$\mathrm{\ddot{o}}$dinger--Foldy equation (4) in the form

\begin{equation}
\label{eq35}
 \left(\partial_{0}+i\widehat{\omega}\right)f(t,\overrightarrow{x})=0; \quad \widehat{\omega} \equiv \sqrt{\overrightarrow{p}^{2} + m^2} =\sqrt { - \Delta + m^2}\geq m>0,
\end{equation}

\noindent and the FW equation in the form

\begin{equation}
\label{eq36}
 \left(\partial_{0}+i\gamma^{0}\widehat{\omega}\right)\phi(t,\overrightarrow{x})=0.
\end{equation}

\noindent We also rewrite the Dirac equation similarly in the form

\begin{equation}
\label{eq37}
 \left(\partial_{0}+i(\overrightarrow{\alpha}\cdot \overrightarrow{p}+\beta m)\right)\psi(t,\overrightarrow{x})=0
\end{equation}

\noindent only for the reasons of analogy and orderliness. Note that the FW transformation between the FW and the Dirac models

\begin{equation}
\label{eq38}
 V^{\pm}\equiv\frac{\pm i\gamma^{l}\partial_{l}+\widehat{\omega}+m}{\sqrt{2\widehat{\omega}(\widehat{\omega}+m)}}
\end{equation}

\noindent is well-defined both for the Hermitian and anti-Hermitian operators.

It is easy to verify that the FW equation (36) follows from the Schr$\mathrm{\ddot{o}}$dinger--Foldy equation (35) 

\begin{equation}
\label{eq39}
 v\left(\partial_{0}+i\widehat{\omega}\right)v=\left(\partial_{0}+i\gamma^{0}\widehat{\omega}\right)\leftrightarrow v\left(\partial_{0}+i\gamma^{0}\widehat{\omega}\right)v=\left(\partial_{0}+i\widehat{\omega}\right)
\end{equation}

\noindent and the general solution of the FW equation (36) follows from the general solution (8) of the Schr$\mathrm{\ddot{o}}$dinger--Foldy equation (35)

\begin{equation}
\label{eq40}
 \phi(t,\overrightarrow{x})=vf(t,\overrightarrow{x})\leftrightarrow f(t,\overrightarrow{x})=v\phi(t,\overrightarrow{x}).
\end{equation}

\noindent Corresponding links between the FW and the Dirac equations are well-known from [1].

Thus, we are able to find the general transformation, which gives relationship directly between the RCQM and the Dirac model

\begin{equation}
\label{eq41}
 W=V^{+}v, \quad W^{-1}=vV^{-}; \quad WW^{-1}=W^{-1}W=1,
\end{equation}

\noindent and to derive the Dirac equation from the RCQM (from the Schr$\mathrm{\ddot{o}}$dinger--Foldy equation)

\begin{equation}
\label{eq42}
 W\left(\partial_{0}+i\widehat{\omega}\right)W^{-1}=\partial_{0}+i(\overrightarrow{\alpha}\cdot \overrightarrow{p}+\beta m),
\end{equation}

\begin{equation}
\label{eq43}
 \psi(t,\overrightarrow{x})=Wf(t,\overrightarrow{x}).
\end{equation}

\noindent The vice versa links also exist as a well-defined non-singular transformations

\begin{equation}
\label{eq44}
 W^{-1} (\partial_{0}+i(\overrightarrow{\alpha}\cdot \overrightarrow{p}+\beta m))W=\partial_{0}+i\widehat{\omega},
\end{equation}

\begin{equation}
\label{eq45}
 f(t,\overrightarrow{x})=W^{-1}\psi(t,\overrightarrow{x}).
\end{equation}

\noindent but are not so interesting for our purposes as the direct transformations (42), (43). The direct transformations derive the Dirac equation from the more elementary model of the same physical reality.

\section{Conclusions}

The model of relativistic canonical quantum mechanics on the level of axiomatic approaches to the quantum field theory is considered. The main intuitive physical principles, reinterpreted on the level of modern physical methodology, mathematically correctly are mapped into the basic assertions (axioms) of the model. The Einstein's principle of relativity is mapped as a requirements of special relativity. The principles of heredity and correspondence of the model with respect to the non-relativistic classical and quantum mechanics are supplemented by the clarifications of external and internal degrees of freedom carriers. The principle of relativity of the model with respect to the means of cognition is realized by the applications of the rigged Hilbert space. The Schwartz test function space $\mathrm{S}^{3,4}$ is shown to be the sufficient to satisfy the requirements of the principle of relativity of the model with respect to the means of cognition. And the fulfilling of calculations in $\mathrm{S}^{3,4}$ does not lead to the loss of generality of the consideration.

It is shown that the algebra of experimentally observed quantities, associated with the Poincar$\mathrm{\acute{e}}$-invariance of the model, is determined by the nine functionally independent operators $\overrightarrow{x}, \overrightarrow{p}, \overrightarrow{s}$, which in the relativistic canonical quantum mechanics model of the doublet have the unambiguous physical sense. It is demonstrated that the application of the stationary complete sets of operators of the experimentally measured physical quantities guarantees the visualization and the completeness of the consideration.

Derivation of the Foldy-Wouthuysen and the Dirac equations from the Schr$\mathrm{\ddot{o}}$dinger-Foldy equation of relativistic canonical quantum mechanics is presented and briefly discussed. We prove that the Dirac equation is the consequence of more elementary model of the same physical reality. The relativistic canonical quantum mechanics is suggested to be such fundamental model of the physical reality. Moreover, it is suggested to be the most fundamental model of the Fermi spin $s=\frac{1}{2}$-doublet.

\textbf{An important assertion is that} \textit{an arbitrary physical and mathematical information, which contains in the model of relativistic canonical quantum mechanics, is translated directly and unambiguously into the information of the same physical content in the field model of the Dirac equation.}

Hence, the Dirac equation is the unambiguous consequence of the relativistic canonical quantum mechanics of the Fermi spin $s=\frac{1}{2}$-doublet (e. g. $e^{-}e^{+}$-doublet). Nevertheless, the model of relativistic canonical quantum mechanics of the Fermi-doublet has evident independent application. 

We pay attention that the model of the relativistic canonical quantum mechanics of the Fermi-doublet does not need the application of the positron negative mass concept [42--45]. It is natural due to the following reasons. It is only the energy, which depends from the mass. And the total energy together with the momentum is associated with the external degrees of freedom, which are common and the same for the particle and antiparticle (for the electron and positron). The difference between $e_{-}$ and $e_{+}$ contains only in internal degrees of freedom such as the spin $\overrightarrow{s}$ and sign of the charge $g=-\gamma^{0}$. Thus, if in the relativistic canonical quantum mechanics the mass of the particle is taken positive then the mass of the antiparticle must be taken positive too.

On the other hand the comprehensive analysis [43] of the Dirac equation for the doublet had led the authors of the article [43] to the concept of the negative mass of the antiparticle. Therefore, our consideration in the last paragraph gives the additional arguments that the Dirac (or associated with it Foldy-Wouthuysen) model is not the quantum-mechanical ones. Furthermore, in the problem of the relativistic hydrogen atom the use of negative-frequency part $\psi^{-}(x)=e^{-i\omega t}\psi(\overrightarrow{x})$ of the spinor $\psi(x)$ in the "role of the quantum-mechanical object" is not a valid. In this case neither $\left|\psi(\overrightarrow{x})\right|^{2}$, nor $\overline{\psi}(\overrightarrow{x})\psi(\overrightarrow{x})$ is the probability distribution density with respect to the eigen values of the Fermi-doublet coordinate operator. It is due to the fact [1] that in the Dirac model the $\overrightarrow{x}$ is not the experimentally observable Fermi-doublet coordinate operator.

The application of the relativistic canonical quantum mechanics can be useful for the analysis of the experimental situation found in [46]. Such analysis is interesting due to the fact that (as it is demonstrated here in the sections 4 and 5) the relativistic canonical quantum mechanics is the most fundamental model of the Fermi-doublet.

Another interesting application of the relativistic canonical quantum mechanics is inspired by the article [47], where the quantum electrodynamics is reformulated in the Foldy-Wouthuysen representation. The author of [47] essentially used the result of the [43] about the negative mass of the antiparticle. Starting from the relativistic canonical quantum mechanics we are able not to appeal to the conception of the antiparticle negative mass.

\vskip 1.cm


\begin{thebibliography}{99}

\bibitem {1} L. Foldy and S. Wouthuysen, Phys. Rev., \textbf{78}, 29 (1950).

\bibitem {2} L. Foldy,  Phys. Rev., \textbf{102}, 568 (1956).

\bibitem {3} L. Foldy,  Phys. Rev., \textbf{122}, 275 (1961).

\bibitem {4} E. Salpeter,  Phys. Rev., \textbf{87}, 328 (1952).

\bibitem {5} W. Lucha and F. Schobert, Phys. Rev. A., \textbf{54}, 3790 (1996).

\bibitem {6} Y. Chargui and A. Trabelsi. Phys. Lett. A, \textbf{377}, 158 (2013).

\bibitem {7} I. Krivsky, V. Simulik, I. Lamer, T. Zajac, arXiv: 1301.6343 [math-ph] 27 Jan 2013.

\bibitem {8} I. Krivsky, V. Simulik, I. Lamer, T. Zajac, TWMS J. App. Eng. Math., \textbf{3}, 62 (2013).

\bibitem{9} I. Krivsky, V. Simulik, T. Zajac, I. Lamer, Derivation of the Dirac and Maxwell equations from the first principles of relativistic canonical quantum mechanics // Proceedings of the 14-th Internat. Conference "Mathematical Methods in Electromagnetic Theory" - 28-30 August 2012, Institute of Radiophysics and Electronics, Kharkiv, Ukraine. 201-204.

\bibitem {10} N. Bogoliubov, A. Logunov, I. Todorov, \textit{Foundations of the axiomatic approach in quantum field theory} (Nauka, Moskow, 1969) (in Russian).

\bibitem {11} V. Simulik, I. Krivsky, Reports of the National Academy of Sciences of Ukraine, \textbf{5}, 82 (2010).

\bibitem {12} I. Krivsky, V. Simulik, Cond. Matt. Phys., \textbf{13}, 43101 (2010).

\bibitem {13} V. Simulik, I. Krivsky, Phys. Lett. A., \textbf{375}, 2479 (2011).

\bibitem {14} V. Simulik, I. Krivsky, I. Lamer, TWMS J. App. Eng. Math., \textbf{3}, 46 (2013).

\bibitem {15} V. Simulik, I. Krivsky, Ukr. J. Phys., \textbf{58}, 523 (2013).

\bibitem{16} P. Dirac, \textit{The principles of quantum mechanics} (Clarendon Press, Oxford, 1958).

\bibitem{17} N. Bogoliubov and D. Shirkov, \textit{Introduction to the theory of quantized fields} (John Wiley and sons, New York, 1980).

\bibitem{18} J. Sakurai, \textit{Advanced quantum mechanics} (Addison--Wesley Pub. Co, New York, 1967).

\bibitem{19} L. Ryder, \textit{Quantum field theory} (2-nd edit., University Press, Cambridge, 1996).

\bibitem{20} J. Keller, \textit{Theory of the electron. A theory of matter from START} (Kluwer Academic Publishers, Dordrecht, 2001).

\bibitem{21}  V. Fock, D. Iwanenko, Z. Phys., \textbf{54}, 798 (1929).

\bibitem{22} V. Fock, Z. Phys., \textbf{57}, 261 (1929).

\bibitem{23} A. Wightman, in \textit{Dispersion relations and elementary particles, edit. C. De Witt and R. Omnes.} (Wiley and Sons, New York, 1960).

\bibitem{24} H. Sallhofer, Z. Naturforsch. A., \textbf{33}, 1378 (1978).

\bibitem{25} H. Sallhofer, Z. Naturforsch. A., \textbf{41}, 468 (1986).

\bibitem{26} S. Strinivasan, E. Sudarshan., J. Phys. A., \textbf{29}, 5181 (1996).

\bibitem{27} L. Lerner, Eur. J. Phys., \textbf{17}, 172 (1996).

\bibitem{28} T. Kubo, I. Ohba, H. Nitta, Phys. Lett. A., \textbf{286}, 227 (2001).

\bibitem{29} H. Cui, arXiv: quant-ph/0102114. 15 Aug 2001, 4 p.

\bibitem{30} Y.J. Ng, H. van Dam, arXiv: hep-th/0211002. 4 Feb 2003, 9 p.

\bibitem{31} M. Calerier, L. Nottale, Electromagnetic Phenomena., \textbf{3(9)}, 70 (2003).

\bibitem{32} M. Evans, Found. Phys. Lett., \textbf{16}, 369 (2003).

\bibitem{33} M. Evans, Found. Phys. Lett., \textbf{17}, 149 (2004).

\bibitem {34} V.M. Simulik, I.Yu. Krivsky, Rep. Math. Phys. \textbf{50}, 315 (2002).

\bibitem {35} V.M. Simulik, I.Yu. Krivsky, Electromagnetic Phenomena., \textbf{3(9)}, 103 (2003).

\bibitem {36} V.M. Simulik, The electron as a system of classical electromagnetic and scalar fields. \textit{In book: What is the electron?} 109-134 \textit{Edited by Simulik V.M.} (Montreal, Apeiron,
2005).

\bibitem{37} von J. Neumann, \textit{Mathematische Grundlagen der Quantenmechanik} (Verlag von Julius Springer, Berlin, 1932).

\bibitem{38} V. Vladimirov, \textit{Methods of the theory of generalized functions} (Taylor and Francis, London, 2002).

\bibitem{39} G. Bluman, S. Anco, \textit{Symmetry and integration methods for differential equations} (Springer, New York, 2002).

\bibitem{40} Elliott, J.P. and Dawber, P.J., (1979), Symmetry in Physics, Vol.1, Macmillian Press, London.

\bibitem{41} Wybourne, B.G., (1974), Classical groups for Physicists, John Wiley and sons, New York.

\bibitem{42} Bondi, H.,  (1957), Negative mass in general relativity, Rev. Mod. Phys., 29, 423-428.

\bibitem{43} Recami, E. and Zino, G., (1976), About new space-time symmetries in relativity and quantum mechanics, Nuovo Cim. A., 33, 205-215.

\bibitem{44} Landis, G., (1991), Comments on negative mass propulsion, Journ. Propulsion and Power, 7, 304-304.

\bibitem{45} Wayne, R., (2012), Symmetry and the other events in time, A proposed identity of negative mass with antimatter, Turk. Journ. Phys., 36, 165-177.

\bibitem{46} Kuellin, W., Gaal, P., Reimann, K., Worner, T., Elsaesser, T., Hey, R., (2010) Coherent ballistic motion of electrons in a periodic potential, Phys. Rev. Lett., 104, 146602(1-4).

\bibitem{47} Neznamov, V.P., (2006), On the theory of interacting fields in the Foldy-Wouthuysen representation, Phys. Part. Nucl., 37, 86-115.


\end{thebibliography}
\end{document}